\documentclass{iopart}
\usepackage{graphicx}
\usepackage{iopams}

\begin{document}
\title[Causal viscous hydrodynamics]{Causal relativistic 
hydrodynamics for viscous fluids\footnote{Work supported by the U.S. Department
of Energy under grant DE-FG02-01ER41190.} 
}
\author{Ulrich Heinz$^{1,2}$ and Huichao Song$^1$}
\address{$^1$Department of Physics, The Ohio State University, Columbus, 
         OH  43210, USA}
\address{$^2$CERN, Physics Department, Theory Division, 
         CH-1211 Geneva 23, Switzerland}

\begin{abstract}
We report on recent results from VISH2+1, a code that solves the
relativistic Israel-Stewart equations for causal viscous hydrodynamics
for heavy-ion collisions with longitudinal boost invariance. We find 
that even ``minimal'' shear viscosity $\eta/s=\hbar/(4\pi)$ leads 
to a large reduction of elliptic flow compared to ideal fluid dynamics.
We explore systematically the sensitivity of this reduction to the 
equation of state, system size, initial conditions, and the microscopic 
relaxation time in different formulations of the Israel-Stewart equations.
\end{abstract}

\section{Introduction}
A good motivation for studying relativistic hydrodynamics for viscous
fluids in order to describe the expansion of the fireballs created in
relativistic heavy-ion collisions is given in Chaudhuri's plenary talk
\cite{Chaud}. We here describe results from systematic studies done
with VISH2+1, our numerical code for solving viscous Israel-Stewart
\cite{Israel:1979wp} hydrodynamics for longitudinally boost-invariant
systems in two transverse space dimensions. We will concentrate on
the most important results and refer the reader to the original 
publications \cite{Heinz:2005bw,Song:2007fn,Song:2007ux,Song:2008si}
for details. We include only the effects of shear viscosity, neglecting
bulk viscosity and heat conduction.

\section{Israel-Stewart formalism}
The Israel-Stewart formalism \cite{Israel:1979wp} resolves problems of 
acausality and numerical instbility of the relativistic Navier-Stokes
equation by allowing the shear viscous pressure components $\pi^{\mu\nu}$ 
of the energy momentum tensor to evolve dynamically, on a short microscopic
time scale $\tau_\pi$, towards their Navier-Stokes limit. Different forms 
of the equations governing the evolution of $\pi^{\mu\nu}$ are being 
discussed in the literature (see references in \cite{Chaud,Song:2008si}).
Although they all converge to the same Navier-Stokes limit as $\tau_\pi
\to0$, their results disagree for finite $\tau_\pi$ (which is necessary for 
stable numerical implementation) \cite{Chaud,Song:2007fn,Song:2007ux,%
Song:2008si,Romatschke:2007mq}. Recent work \cite{Song:2008si} has shown 
that the weakest sensitivity of physical observables to the "regulator" 
$\tau_\pi$ is obtained by employing the so-called "full" Israel-Stewart 
equations used in \cite{Song:2008si,Romatschke:2007mq,Muronga:2003ta} 
whereas "simplified" versions advocated in \cite{Israel:1979wp,Heinz:2005bw} 
that are obtained by dropping certain terms in the "full" equations and were 
used in \cite{Chaud,Song:2007fn,Song:2007ux,Baier:2006gy} give results that 
depend strongly on the (presently poorly constrained) value for $\tau_\pi$. 
As shown in \cite{Heinz:2005bw} at the end of section IV, the extra terms 
in the "full I-S" formulation cause a reduction of the effective kinetic 
relaxation time for $\pi^{\mu\nu}$, especially in rapidly expanding systems. 
This keeps the viscous pressure smaller than in the "simplified I-S" 
formalism, holding it always close to its Navier-Stokes limit\footnote{We 
thank D. Rischke and U. Wiedemann for an illuminating discussion on this 
point.} and thereby eliminating large excursions of $\pi^{\mu\nu}$ that 
increase viscous entropy production and lead to a stronger viscous reduction 
of the elliptic flow as $\tau_\pi$ is increased \cite{Song:2008si}. The
\"Ottinger-Grmela framework used in \cite{Dusling:2007gi} appears to give
very similar results as the "full I-S" one\footnote{It has recently been 
checked that VISH2+1 with "full I-S" equations and the codes developed by 
P. \& U.~Romatschke \cite{Romatschke:2007mq} and Dusling \& Teaney 
\cite{Dusling:2007gi} give, for identical initial conditions and equations 
of state, the same time evolution for the momentum anisotropy in non-central
Au+Au collisions at RHIC energies (up to small numerical errors). We thank 
K. Dusling for providing us with his results for comparison, and 
P. Romatschke for help with using their code.} \cite{private}.

\section{Cooling rates for ideal and viscous heavy-ion fireballs}
\begin{figure}[htb]
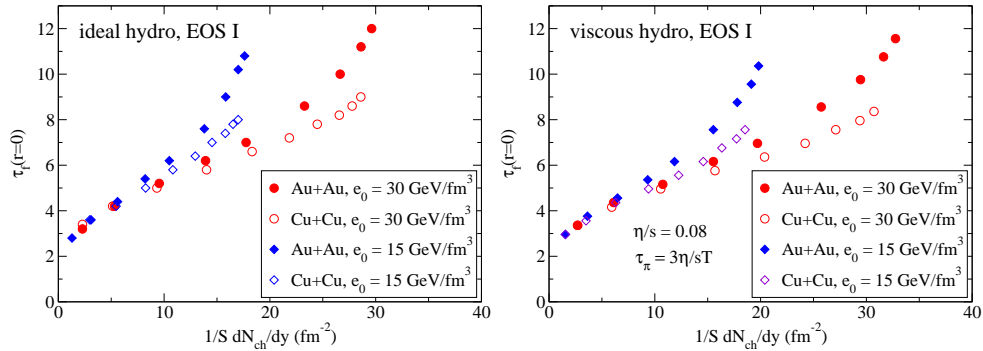

  \begin{center}
 \includegraphics[width=0.49\linewidth,clip=]{3tf_Id_EOSI.eps} 
 \includegraphics[width=0.49\linewidth,clip=]{3tf_Visfull_EOSI.eps}
\vspace*{-2mm}
  \end{center} 
  \caption{\label{F1}(Color online)
  Central freeze-out times for Au+Au and Cu+Cu collisions with
  differential initial peak energy densities $e_0\equiv e(0,0;b{=}0)$
  as a function of final charged hadron multiplicity density.
  Points belonging to one series correspond to different impact
  parameters, with central (peripheral) collisions corresponding
  to large (small) multiplicity densities. Left: ideal fluid dynamics.
  Right: viscous fluid dynamics with $\eta/s=0.08$ and $\tau_\pi=
  3\eta/sT$, using the full I-S equations.
  } 
 \end{figure}

In the early expansion stage of a heavy-ion collision, shear viscosity 
leads to a reduction of the longitudinal and an increase of the transverse 
pressure. The reduced longitudinal pressure decreases the work done by
longitudinal expansion, thereby reducing the initial cooling rate of the
fireball. This leads to a somewhat increased lifetime of the quark-gluon
plasma phase. The increased transverse pressure, on the other hand, causes
larger transverse acceleration and stronger radial flow of the matter than 
in ideal fluid dynamics. Due to the larger transverse flow, the center of 
the viscous fireball cools more quickly during the late stages than an
ideal fluid, thereby slightly reducing the total fireball lifetime until
freeze-out for central and near-central collisions (see Fig.~\ref{F1}).
The viscous fireballs created in peripheral collisions don't live long 
enough for this mechanism to manifest itself; they live longer than their
ideal counterparts, due to the decreased initial cooling rate arising from
the smaller longitudinal pressure (left points in Fig.~\ref{F1}). 
 
%
%

For an equation of state (EOS) with a first-order quark-hadron phase 
transition, the mixed phase matter is free of pressure gradients and 
hence not accelerated. Velocity gradients in the mixed phase generate,
however, viscous pressure components whose gradients continue to accelerate 
the viscous fluid even while it passes through the mixed phase where the
thermal pressure gradients vanish. As a result, the viscous fluid spends 
less time in the mixed phase than the ideal one. 

In ideal hydrodynamics, a first-order phase transition with a mixed phase
(MP) generates large velocity gradients near the QGP-MP and MP-HG (HG =
hadron gas) interfaces. In viscous fluids such gradients generate viscous
pressures that act against building up large velocity gradients. In 
consequence, all prominent structures in ideal hydrodynamics that arise
from discontinuities in the speed of sound in ideal hydrodynamics are 
washed out by viscous effects. Shear viscosity thus effectively turns a
first-order phase transition into a smooth cross-over transition.

\section{Momentum anisotropy in non-central collisions}
In viscous hydrodynamics the total momentum anisotropy 
$\varepsilon_p=\frac{\langle T^{xx}{-}T^{yy}\rangle}
{\langle T^{xx}{+}T^{yy}\rangle}$ has two contributions. The first
arises (as in ideal fluids) from anisotropies in the collective
flow velocity. The second is due to anisotropies of the viscous
shear pressure tensor which contribute a momentum anisotropy even
in the local rest frame of the fluid. In heavy-ion collisions, these
two contributions are found to be of opposite signs: the viscous
pressure effects reduce the total omentum anisotropy. In the early
stages of the expansion, the collective flow anisotropies evolve very
similarly in viscous and ideal fluids; however, due to the large
longitudinal expansion rate $\sim 1/\tau$, the viscous pressure effects
are large at early times, leading to a strong reduction of the total
momentum anisotropy. At late times, the viscous pressure effects are
dramatically reduced, but by then their time-integrated effect has led to 
a slower build-up of collective flow anisotropies compared to ideal fluids.
So the total momentum anisotropy at late times is again much smaller for
the viscous fluid, even though the negative viscous pressure contributions
to this anisotropy are now almost neglegible. Overall, shear viscous effects
are seen to cause during all evolution stages a significant reduction of the 
momentum anisotropy compared to ideal fluids. 

\section{Spectra and elliptic flow}
The larger radial flow caused by the viscous increase of the transverse
pressure gradients reflects itself through flatter transverse momentum
spectra for the finally emitted hadrons. To obtain similar spectral slopes
in viscous and ideal hydrodynamics thus requires a retuning of initial
conditions and/or freeze-out conditions. Starting the viscous expansion
later with lower initial energy densities works \cite{Baier:2006gy,%
Romatschke:2007mq}, but this neglects transverse flow generated during
the earlier pre-equilibrium stages. A consistent retuning of initial 
conditions that is compatible with dynamical models for the 
pre-hydrodynamic stage remains an open issue.

The smaller momentum anisotropies of viscous fireballs discussed above
manifest themselves in a reduction of the elliptic flow coefficients extracted 
from the final hadron spectra. Initial elliptic flow results from our group 
\cite{Song:2007fn,Song:2007ux} and Romatschke \cite{Romatschke:2007mq} seemed 
to show large disagreements: where we saw for $b=7$\,fm Cu+Cu collisions at 
$p_T=2$\,GeV/$c$ a reduction of pion $v_2$ by about 63\% from the ideal fluid 
value, Romatschke saw for $b=7$\,fm Au+Au collisions only a 25\% reduction. 
These differences have been understood and resolved in \cite{Song:2008si}. 
The largest part of the discrepancy results
from the different system sizes: going from Cu+Cu to Au+Au at the same
impact parameter cuts the viscous reduction of $v_2$ almost in half,
to 39\%. Viscous effects are much more dramatic for small 
collision fireballs than for larger ones, due to larger transverse velocity 
{\em gradients}. Accounting for different EOS used in \cite{Song:2007fn,%
Song:2007ux} and \cite{Romatschke:2007mq} removes most of the remaining 
difference: replacing SM-EOS~Q (with a first order phase transition) by 
EOS~L (with a smooth cross-over fitted to lattice QCD data) reduces the 
viscous suppression of $v_2$ by about another third, to 28\%. A final 5-10\% 
relative reduction of the suppression results from 
replacing the "simplified I-S" treatment of \cite{Song:2007fn,Song:2007ux} 
with the "full I-S" formalism used in \cite{Romatschke:2007mq}. The combined 
effect of these changes is a viscous suppression of $v_2$ of 25\% in 
$b=7$\,fm Au+Au at $p_T=2$\,GeV/$c$, which agrees with 
\cite{Romatschke:2007mq}.

\section{Multiplicity scaling of elliptc flow and viscous entropy 
production}
Voloshin and collaborators pointed out (see \cite{v2scaling}
and references therein) that the eccentricity-scaled elliptic flow
$v_2/\varepsilon_x$ scales with the final hadron multiplicity per unit 
overlap area, $(1/S)(dN_\mathrm{ch}/dy)$. All dependence on system size, 
beam energy, and impact parameter can apparently be subsumed in this variable. 
The empirical scaling function rises monotonically with multiplicity 
density, approaching the ideal fluid prediction at high multiplicities
but remaining well below it at low multiplicities.
 
We investigated this "multiplicity scaling" for ideal and viscous 
hydrodynamics, for both $v_2/\varepsilon_x$ and the viscous entropy production 
fraction $\Delta{\cal S}/{\cal S}_0$ \cite{Song:2008si}. Even in the ideal 
fluid case we found weak scaling violations that can be traced back to the 
freeze-out condition which introduces an external scale that affects the 
fireball lifetime (see Fig.~\ref{F1}). Perfect multiplicity scaling of 
$v_2/\varepsilon_x$ in ideal fluid dynamics \cite{Bhalerao:2005mm} can 
only be achieved only if decoupling happens late enough that the momentum 
anisotropy can fully saturate. In \cite{Song:2008si} we found this to be 
the case for very high collision energies.

Scale breaking is more pronounced in viscous hydrodynamics and increases 
with $\eta/s$. Experimentally seen weak scale breaking effects appear to 
be qualitatively consistent with viscous hydrodynamic predictions 
\cite{Song:2008si} but require more precise data for confirmation.

Approximate multiplicity scaling is also found for the viscous entropy 
production fraction. $\Delta{\cal S}/{\cal S}_0$ depends strongly on the 
hydrodynamic starting time $\tau_0$. For $\tau_0=0.6$\,fm/$c$
and EOS~L we find entropy production fractions ranging from about 10\%
at high multiplicity densities to about 12-13\% at low multiplicities;
again smaller systems or more peripheral collisions exhibit stronger viscous 
effects.



\section*{References}
\begin{thebibliography}{99}

\bibitem{Chaud}
  Chaudhuri A K 2008 these proceedings.
  
\bibitem{Israel:1979wp}
  Israel W 1976 {\it Annals Phys.} {\bf 100} 310;
  Israel W and Stewart J M 1979
  {\it ibid.} 
  {\bf 118} 341  
  
\bibitem{Heinz:2005bw}
  Heinz U, Song H and Chaudhuri A K 2006
  {\it Phys. Rev.}  C {\bf 73} 034904.

\bibitem{Song:2007fn}
  Song H and Heinz U 2008
  {\it Phys.\ Lett.}  B {\bf 658} 279.

\bibitem{Song:2007ux}
  Song H and Heinz U 2008
  {\it Phys. Rev.} C, in press 
  ({\it preprint} arXiv:0712.3715 [nucl-th]).

\bibitem{Song:2008si}
  Song H and Heinz U 2008
  {\it preprint} arXiv:0805.1756 [nucl-th].

\bibitem{Romatschke:2007mq}
  Romatschke P and Romatschke U 2007
  {\it Phys.\ Rev.\ Lett.} {\bf 99} 172301;
  Luzum M and Romatschke P 2008
  {\it preprint} arXiv:0804.4015 [nucl-th].

\bibitem{Muronga:2003ta}
  Muronga A 2004
  {\it Phys.\ Rev.}  C {\bf 69} 034903.

\bibitem{Baier:2006gy}
  Baier R and Romatschke P 2007
  {\it Eur.\ Phys.\ J.}  C {\bf 51} 677; 
  Romatschke P 2007
  {\it ibid.}
  {\bf 52} 203. 

\bibitem{Dusling:2007gi}
  Dusling K and Teaney D 2008
  {\it Phys.\ Rev.}  C {\bf 77} 034905.

\bibitem{private}
  Dusling K and Teaney D 2008 private communication.

\bibitem{v2scaling}
  Voloshin S A 2006 {\it AIP Conf. Proc.} {\bf 870} 691;
  Voloshin S A 2007 {\it J. Phys. G: Nucl. Part. Phys.} {\bf 34} S883.
   
\bibitem{Bhalerao:2005mm}
  Bhalerao R S, Blaizot J P, Borghini N and Ollitrault J Y 2005
  {\it Phys.\ Lett.}  B {\bf 627} 49.

\end{thebibliography}
\end{document}